\documentclass{article}
\usepackage{emulateapj,psfig}

\def\simlt{\mathrel{\spose{\lower 3pt\hbox{$\mathchar"218$}}
     \raise 2.0pt\hbox{$\mathchar"13C$}}}
\def\simgt{\mathrel{\spose{\lower 3pt\hbox{$\mathchar"218$}}
'     \raise 2.0pt\hbox{$\mathchar"13E$}}}
\def\gsim{ \lower .75ex \hbox{$\sim$} \llap{\raise .27ex \hbox{$>$}} }
\def\lsim{ \lower .75ex \hbox{$\sim$} \llap{\raise .27ex \hbox{$<$}} }

\accepted{}

\long\def\***#1{{\scshape ***#1***}}
\def\arraystretch{1.25}



\lefthead{Macci\`o et al.}
\righthead{Mass of clusters in simulations}

\def\be{\begin{equation}}
\def\ee{\end{equation}}
\def\bea{\begin{eqnarray}}
\def\eea{\end{eqnarray}}
\def\mincir{\raise -2.truept\hbox{\rlap{\hbox{$\sim$}}\raise5.truept
\hbox{$<$}\ }}
\def\magcir{\raise -4.truept\hbox{\rlap{\hbox{$\sim$}}\raise5.truept
\hbox{$>$}\ }}
\def\lsim{\raise -2.truept\hbox{\rlap{\hbox{$\sim$}}\raise5.truept
\hbox{$<$}\ }}
\def\gsim{\raise -4.truept\hbox{\rlap{\hbox{$\sim$}}\raise5.truept
\hbox{$>$}\ }}

\begin{document}

\title{Mass of clusters in simulations}

\author{Andrea V. Macci\`o\altaffilmark{1,2}, Giuseppe Murante\altaffilmark{3} \& 
Silvio A. Bonometto\altaffilmark{1,2}}

\altaffiltext{1}{Department of Physics G. Occhialini, Universit\`a di 
Milano--Bicocca, Piazza della Scienza 3, I~20126 Milano (Italy)}

\altaffiltext{2}{I.N.F.N., Via Celoria 16, I20133 Milano (Italy)}

\altaffiltext{3}{I.N.A.F., Osservatorio Astronomico di Torino -- Torino (Italy)}

\begin{abstract}

We show that dark matter halos, in n--body simulations, have a 
{\it boundary layer} (BL), which neatly separates dynamically bound 
mass from unbound materials. 
We define $T(r)$ and $W(r)$ as the { differential} kinetic and
potential energy of halos and evaluate them in { spherical 
shells}. We notice that, in simulated halos, such differential
quantities fulfill the following properties:
(i) The differential virial ratio ${\cal R}=-2T/W$ has at least one 
{\it persistent} (resolution independent) minimum ${\bar r}$, such that,
close to ${\bar r}$, (ii) the function  $w=-d\log W/d\log r$ has
a maximum, while (iii) the relation ${\cal R}({\bar r})\simeq w({\bar r})$
holds. BL's are set where these three properties are fulfilled, in
halos found in simulations of TCDM and $\Lambda $CDM
models, run {\it ad--hoc}, using the ART and GADGET codes;
their presence is confirmed in larger simulations of the same models,
with a lower level of resolution. Here we find that $\sim 97\, \%$
of the $\sim 300$ clusters (per model) we have with $M>4.2 \cdot 
10^{14} h^{-1} M_\odot$, owns a BL. Those clusters which appear not 
to have a BL are seen to be undergoing major merging processes and to 
grossly violate spherical symmetry.
The radius ${\bar r} \equiv r_c$ has significant properties. 
First of all, the mass $M_c$ it encloses almost coincides with the
mass $M_{dyn}$, evaluated from the velocities of all particles within
$r_c$, according to the virial theorem. Also, materials at $r
> r_c$ are shown not be in virial equilibrium. Using $r_c$ we can 
then determine
an individual density contrast $\Delta_c$ for each virialized halo, 
that we compare with the "virial" density contrast $\Delta_v 
\simeq 178 \, \Omega_m^{\, 0.45}$ (where $\Omega_m$ is the matter
density parameter)
obtained assuming a spherically symmetric and unperturbed fluctuation growth.
As expected, for each mass scale, $\Delta_v$ is within the range of 
$\Delta_c$'s. However the spread in $\Delta_c$ is wide, while the 
average $\Delta_c$ is $\sim 25\, \%$ smaller than the corresponding 
$\Delta_v$. 
We argue that the matching of properties derived under the assumption
of spherical symmetry must be a consequence of an approximate sphericity,
after violent relaxation destroyed features related to ellipsoidal
non--linear growth. On the contrary, the spread of final $\Delta_c$
is an imprint of the different initial 3--D geometries of fluctuations 
and of the variable environment during their collapse, as suggested by
a comparison of our results with Sheth \& Tormen analysis.

\end{abstract}


\section{Introduction}
The inner region of dark matter halos in n--body simulations 
is certainly virialized. A precise detection of the boundaries
of such a region, however, is a long--lasting and not yet completely 
solved problem. If we try just to identify a ``collapsed'' region, 
$i.e.$ a region detached from the cosmological expansion, the task 
is easier (see, e.g., Monaco \& Murante 1999 and references therein). 
However, in general, a collapsed region is not yet in virial equilibrium.
The point is that, in a cosmological context, each structure is embedded
in a more or less homogeneous background, and we pass gradually from 
the inner halo to external materials. 
The aim of this work is to show a precise criterion able to
find virialized regions in dark matter halos. This criterion
will be tested on simulations and found to be more efficient
than previously adopted rules.

As a matter of fact, several global properties of 
galaxy clusters, such as mass ($M$), 
radius ($r$) and, also, temperature ($T_X$) or luminosity ($L_X$) are 
strictly linked to the achievement of equilibrium.
Analytical predictions on such global quantities were performed
assuming that structures evolve from uniform, spherically symmetric
fluctuations, preserve such symmetries during their growth and feel
no influence from their environment (Gott \& Rees 1975, Peebles 1980). 
However, even under such simplifying conditions, 
a cosmological constant is a significant complication.
Lahav et al (1991) studied the growth of uniform spherical fluctuations
within such context. They also outlined that the most direct way
through which observations can distinguish among flat models with 
different $\Omega_m$ (matter density parameter) is comparing
cluster abundances at low and high redshift.
The dynamics of the spherically symmetric 
growth of a uniform density fluctuation,
in a model with $\Omega_m + \Omega_\Lambda =1$ (where $\Omega_\Lambda$
is the 
vacuum density parameter), was further explored by Eke et al (1996). 
Brian and Norman (1998) gave a polynomial expression for
the expected density contrast of virialized clusters, which reads 
$\Delta_v = A+Bx+Cx^2$, with $x=\Omega_m-1$, and $A=18\pi^2 \simeq 178$,
$B=82$, $C=-39$, and holds at any redshift $z$; this relation is well
approximated by the expression:
$$
\Delta_v = A \Omega_m^{\, 0.45} ~.
\eqno (1.1)
$$
Under the same assumptions, using a Press \& Schechter (1974) 
approximation, one can predict the expected mass function of clusters. 

The impact of this approximation on the estimate of cluster global 
quantities could be appreciated only through numerical simulations. 
Much work was therefore devoted to this aim (see, e.g., Lacey \& 
Cole, 1993,1994; Cole \& Lacey 1996; Carlberg et al., 1996; Eke et 
al. 1996; Eke et al. 1998; Brian \& Norman, 1998; Gardini et al. 1999). 
However, in the very analysis of the numerical work, the above approximations
were often implicit. An example is the identification of dark halos 
based on spherical overdensity (SO) algorithms, with reference to the 
density contrast $\Delta_v$ given in eq.~(1.1). In this work we shall
also use an SO algorithm (described below) as a selection criterion. 
However, using eq.~(1.1), to work out values for the cluster radius ($r_v$) or
mass ($M_v$), implies a bias. The values (1.1) are obtained under precise
restrictions and numerical work should also verify their impact.
In a sense, the problem is even more severe when the values (1.1)
are used to define virialized halos without requiring explicitly
that they are spherical.

In this work we plan to find a rule, suitable for defining
virialized systems, taking into account that a variety of growth
histories led to different final features of virialized structures.
A number of recent papers (Sheth \& Tormen (1999, 2002), Sheth, Mo 
\& Tormen, 2001) had a similar motivation. They tried to study the
growth of primeval fluctuations taking into account the effect of an 
ellipsoidal collapse. Then, in order to avoid having the collapse 
on some axis going to zero, they set up a recipe to freeze it 
out once it has shrunk by some critical factor.  This freeze--out radius 
is explicitly chosen so that the density contrast at virialization for
the whole halo is again $\Delta_v$ as given in eq.~(1.1), 
$i.e.$ in spherical growth models. 
In contrast, in this work the use of $\Delta_v$ to define
virialized systems is completely overcome. The point is that
the detailed growth history of individual 
halos, in which different ellipticities and velocity anisotropies 
played an important role, is not expected to leave major traces on final
halo shapes after violent relaxation has occurred during the 
virialization process (Navarro, Frenk \& White 1997). 
On the other hand, it may result in a significant spread of the 
equilibrium density contrasts $\Delta_c$ of individual halos. 
Therefore, when the virial radius of relaxed objects is sought,
deviations from spherical symmetry and velocity anisotropies may not
be as important as deviations of $\Delta_c$ from $\Delta_v$.

We expect that the value $\Delta_v$, given by eq.~(1.1),
is within the range of the actual $\Delta_c$ values, but we think
that it is important to first test how extended this range is,
and how $\Delta_v$ is set with respect to actual values.
Furthermore, a different $\Delta_c$ implies a different radius and
mass ($r_c$ and $M_c$) for each object. Accordingly, we expect that
some individual object masses increase and others decrease.
We will explore the extent of these changes on galaxy clusters
in numerical simulations of critical CDM models
with and without a cosmological constant.
Such changes may have an impact on the mass function (see section 5 and 
Fig.~13 and 14), while some scaling relations might be 
significantly altered, for instance the relation between $M$ and $T_X$.

In order to implement our program, we need a simple and effective rule
to identify virialized material in clusters. 
In trying to find such rule we discovered a precise regularity in the
transition region between material belonging to clusters
and the surrounding medium. In fact, the volume occupied by virialized 
materials is
confined by a {\it boundary layer} (BL), whose depth $\Delta r$ ranges around
50--100$\, h^{-1}$kpc, and whose properties fulfill
precise analytical prescriptions. 
Such a BL has been found in $\sim 97\, \%$ of the clusters in
our simulated models and is one of the findings of this work.

The BL is not a physical confinement barrier, and particles can travel
outwards and inwards through it. On the contrary, it confines a
volume, where 
material fulfills precise equilibrium conditions,
which cease to hold outside it. Inside the BL itself, 
the formal condition for virial equilibrium of an isolated system
holds, as pressure forces at its upper and lower borders cancel 
each other; on the BL, kinetic energy has a minimum with respect to
potential energy, and matter density is rapidly decreasing.
These conditions follow from a precise analytical requirement
($Rw$ requirement) derived under the assumption of spherical symmetry.
This $Rw$ requirement will be only approximately satisfied by halos
in simulations, where the halo geometries apparently 
violate spherical symmetry.
We show, however, that normal deviations from a spherical shape in DM halos
do not prevent them from having a BL.

We explicitly note that we do not {\it force} the BL to exist, 
but we apply the $Rw$ requirement and {\it find} it.
For this reason, finding the BL is different 
from finding a given spherical overdensity; the latter always exists
when average densities of halos decrease with increasing radii,
and this does not imply even an approximately spherical symmetry.
On the contrary, we could have bound and relaxed regions which
are not confined by a BL: our analysis of numerical simulations shows
that this is not the case. There
is also a major difference between the $Rw$ requirement we are
defining here, and other prescriptions, such as SKID, which 
defines a halo as the set of all particles {\it gravitationally bound} 
to the halo itself. In fact,
bound particles are not guaranteed to belong to virialized zones and 
may belong to regions which are bound but still unrelaxed. 

In this paper we shall first discuss when the BL should exist and 
introduce the $Rw$ requirement used to detect it.
To have significant statistics based on this rule
and to confirm its validity
requires simulations performed with a large dynamic 
range, as can only be done by parallel computing.

Finding and situating BL's allows a better evaluation of individual 
cluster masses, radii and density contrasts, as well as an analysis of their 
possible dependence on various physical parameters. The study of the structure 
of dark matter halos in numerical simulations can also be put on a sounder 
footing. In previous analyses (see, e.g.,  Cole \& Lacey 1996) the $r$
dependence of the integral virial ratio $-2T(<r)/W(<r)$ was discussed 
and found not to be ``useful for defining the boundaries of the 
virialized region''. One of the findings of our analysis is precisely
the possibility of detecting such a boundary, using the virial ratio
for shells instead of spheres. 
Fig.~1 compares integral and differential virial ratios,
showing how much more information appears to be contained
in the latter (see Sec.~3 for details on how ${\cal R}(r)$
points are calculated). Previous analyses have also stressed
halo properties apparently depending ``on how groups are identified'' 
(Sheth et al., 2001). The existence
of a border, with precise physical properties, secluding
cluster material lets one overcome any such ambiguity.

It may be also important to remember that according to the definitions
given above, in this work $\Delta_v$, $r_v$, $M_v$ are quantities 
worked out starting from the fixed density contrast given by (1.1), 
while $\Delta_c$, $r_c$, $M_c$ are the same quantities when worked 
out from the setting of the BL around each cluster.
Let us also point out that 
CHb
the procedure discussed in this work is not the basis for an
CHe
alternative cluster--finding algorithm, 
such as, e.g., Friends-of-Friends, SKID or SO. 
On the contrary, our analysis assumes that cluster 
locations are given. Here we give a different prescription for defining
the physical dimension of the cluster, not its position.

The plan of the paper is as follows. In section 2 we give suitable
information on the simulations used, which were partially run
{\it ad--hoc}, for this work. In section 3 and 4 we define the
boundary layer (BL) and discuss how it can be found in simulated
clusters. In section 5 we show the results of our work, and
in section 6 we discuss them and some future perspectives.

\section{Models and n--body simulations}

We use two main sets of simulations, performed with different
codes: the parallel AP3M N--body code described in Gardini et al. 
(1999), the parallel PM ART code developed by Kratshov \& Klypin (1997),
and the parallel tree--code GADGET (Springel et al. 2001).

The code of Gardini et al. (1999) was developed from the serial public 
AP3M code of Couchman (1991), extending it to different cosmological
models and allowing suitable flexibility in particle masses. Using
this code, two simulations were performed. The first of them, dealing 
with a ``tilted'' Einstein-de Sitter model (hereafter TCDM), is the
same already considered by Gardini et al. (1999); the normalization of the run
was, however, rescaled to yield $\sigma_8=0.55$ at the final step, 
instead of $\sigma_8=0.61$, so that the final abundance of rich clusters 
is the same for both models at the final epoch
(as ususal, $\sigma_8$ is the m.s. density fluctuation on the scale
of $8\, h^{-1}$Mpc, $h$ being the Hubble parameter in units of 100 km/s/Mpc). 
The latter simulation 
dealt with a $\Lambda$CDM model, i.e. a flat CDM universe with non--zero
cosmological constant
with $\sigma_8 = 1.08$.
Owing to recent observational results (see, e.g., Schuecker et al, 2002),
these normalizations are both rather high, as a reasonable value for
$\sigma_8$ in $\Lambda$CDM models is $\sim 0.75$; however, this 
has no impact when studying inner cluster dynamical properties.

The above simulations (denoted A and B, respectively) are performed 
in cubes with sides of 360$\, h^{-1}$Mpc.
CDM+baryons are represented by $180^3$ particles,
whose individual mass is $2.22 \cdot 10^{12} h^{-1} M_\odot$ for TCDM
and $0.777 \cdot 10^{12} h^{-1} M_\odot$ for $\Lambda$CDM.
We use a 256$^3$ grid to compute the FFT's needed to evaluate the long 
range contribution to the force (PM) and we allow for mesh refinement
where the particle density attains or exceeds $\sim 30$ times the mean value.
The starting redshifts are $z_{in}=10$ for TCDM and $z_{in}=20$ for 
$\Lambda$CDM. The particle sampling of the density field is obtained by
applying the Zel'dovich approximation (Zel'dovich 1970; Doroshkevich 
et al. 1980) starting from a regular grid. We adopt the same random 
phases in both A and B.

The number of steps were 1000 equal $p$--time steps (the time
parameter is $p \propto a^{2/3}$, where $a$ is the expansion factor).
The comoving force resolution, given by the softening length,
is $\eta \simeq 112\, h^{-1}$kpc, yielding a Plummer equivalent
softening $\epsilon_{pl} \simeq 40.6\, h^{-1}$kpc. 
These simulations were described in more detail in Gardini 
et al. (1999), and were also used in Macci\`o et al. (2002). 
The parameters of the models are reported in detail in Table 1.

In order to see how resolution affects the results of the $Rw$ requirement,
we performed further simulations with the ART code (courtesy of A. Klypin). 
This allows us to select regions inhabited by clusters and to re--run 
them with increasing particle mass resolution: from $7.7 \times 10^{11} 
\Omega_m h^{-1} M_\odot$ to $1.2 \times 10^{10} \Omega_m h^{-1} M_\odot$. 

The ART code (Adaptive Refinement Tree: Kravtsov et al. 1997, 
Knebe et al. 2000) starts with a uniform grid, which covers the whole 
computational box. This grid defines the lowest (zeroth) level of 
resolution of the simulation. The standard Particles-Mesh algorithms 
are used to compute the density and gravitational potential on the 
zeroth-level mesh.  The code then reaches high force resolution by 
refining all high density regions using an automated refinement algorithm.  
The refinements are recursive: the refined regions can also be refined, 
each subsequent refinement having half of the previous level's cell size.
This creates a hierarchy of refinement meshes of different resolution, 
size, and geometry covering regions of interest. Because each individual
cubic cell can be refined, the shape of the refinement mesh can be 
arbitrary and effectively match the geometry of the region of interest.
This algorithm is well suited for simulations of a selected region 
within a large computational box.

The criterion for refinement is the local density of particles: if the
number of particles in a mesh cell (as  estimated by the Cloud-In-Cell
method)  exceeds  the   level  $n_{\rm thresh}$,  the   cell  is split
(``refined'')  into  8   cells of   the   next refinement  level.  The
refinement threshold may depend on the refinement level. The code uses
the   expansion  parameter  $a$ as  the    time variable.   Besides
spatial refinements, during the integration  time refinements are also 
performed.

The ART code can handle particles of different masses. This lets us
increase the mass (and correspondingly the force)
resolution of regions centered around the highest mass clusters.  The
multiple mass resolution is implemented in the following way.  We
first set up a realization of the initial spectrum of perturbations,
so that initial conditions for our largest number ($512^3$) of
particles can be generated in the simulation box.  Coordinates and
velocities of all the particles are then calculated using all waves
ranging from the fundamental mode $k=2\pi/L$ to the Nyquist frequency
$k=2\pi/L\times N^{1/3}/2$, where $L$ is the box size and $N$ is the
number of particles in the simulation.  
We obtained lower resolution regions by merging
high resolution particles into particles of larger mass where needed. 
This process can be 
repeated  to get still lower resolution.  The larger mass (merged)
particles have  velocities and displacements equal to the average of the
velocities and displacements of the smaller-mass particles. 
Using this technique we firstly generated initial conditions for 
low--resolution simulations. These simulations were run from their 
initial redshift to $z=0$ and they were used to locate the Lagrangian 
regions forming the most massive clusters. Then, we were able to 
obtain initial conditions at high resolution (both in force and mass) 
in the selected Lagrangian zones, with the same random Fourier
phases. Such zones were surrounded by lower and lower resolution
shells taking into account the tidal field which acts on them. 
Finally, these ICs were used to run the high--resolution cluster simulations.

The simulations employed here were performed using $128^3$ zeroth-level
grid in a computational box of $180h^{-1}{\rm Mpc}$.  The threshold for cell
refinement (see above) was low on the zeroth level: $n_{\rm
thresh}(0)=2$.  Thus, every zeroth-level cell containing two or more
particles was refined.  
The threshold was higher on deeper levels of refinement: $n_{\rm
thresh}=3$ and $n_{\rm thresh}=4$ for the first level and higher
levels, respectively.

For the low resolution runs the step in the expansion parameter was
chosen to be $\Delta a_0=2\times 10^{-3}$ on the zeroth level of
resolution. This gives about 500 steps for particles located in
the zeroth level for an entire run to $z=0$ and 128.000 for particles
at the highest level of resolution.

Using the ART code we performed two simulations: a $\Lambda$CDM model
and a TCDM model, that we will indicate as C and D, respectively.
These models are similar, although not identical, to those of Gardini 
et al. 1999; their parameters are listed in Table 2. We identified 
the clusters in these zero--level simulations and selected the 6 
largest mass clusters for each cosmological model. They were re--run 
with increased mass (and force) resolution, using the method 
described above. For every selected cluster we have three runs 
with resolutions
\footnote{ Since our averaging procedure corresponds to using different
mesh sizes when generating ICs for a simulation, we will refer to 
the size of the higher resolution mesh used in each simulation IC
when quoting resolutions.  }
of $128^3$  (particles mass $M_p=7.7 \times 10^{11} 
\Omega_m\, h^{-1} M_\odot$), $256^3$ ($M_p=9.6 \times 10^{10} 
\Omega_m\, h^{-1} M_\odot$) and $512^3$ ($M_p=1.2 \times 10^{10} 
\Omega_m\, h^{-1} M_\odot$) with more than 350.000 particles 
within a radius of 2.0 Mpc h$^{-1}$.

We then run 3 out of the 6 high resolution $\Lambda$CDM clusters 
with the public parallel tree-code GADGET (Springel et al. 2001), 
to verify that our results do not depend upon the peculiarities 
of one N-body code. One of the clusters was magnified with both 
ART and GADGET. The initial conditions were set as described above. 
We used a Plummer-equivalent softening length $\epsilon=10.98$; 
this is also the linear dimension or our highest--resolution ART cell. 
We chose the time--step criterion 3, based on the local dynamical time, 
with a tolerance of the integration error $E_{int}=0.2$, which gave us 
more than 100000 time steps from $z_{in}$ to $z=0$. The criterion for 
opening the tree--cells is based on the absolute truncation error in 
the multipole expansion, with a tolerance on the error on the force 
of $E_{F}=0.02$ (see Springel et al. 2001 and the code user manual
for more details).

In all of our numerical simulations, clusters were identified using 
a SO algorithm. As a first step, candidate clusters are located by 
a FoF algorithm, with linking length $\lambda = \phi \times d$ 
(here $d$ is the average particle separation), keeping groups with 
more than $N_f$ particles. We then perform two operations: 
(i) we find the point, $C_W$, where the gravitational potential, due
to the group of particle, is minimum; (ii) we determine the radius,
$\bar r$, of a sphere centered in $C_W$ where the density contrast is
$\Delta_{v}$. Using all particles in this sphere we perform again the
operations (i) and (ii). The procedure is iterated until we converge
onto a stable particle set. The set is discarded if, at some stage, we 
have less than $N_f$ particles. If a particle is a potential member of 
two groups it is assigned to the more massive one. (Gardini et al. 1999 
also describe this SO algorithm in more detail and compare it with other 
group identification algorithms, e.g. Governato et al. 1999).
In this work we set $\phi = 0.2$ and take $N_f$ so to have
a mass threshold $3.0\times 10^{13}\, h^{-1}M_\odot$. Above this mass
threshold there are $\sim 10000$ {\it halos }in the simulations A,B and
$\sim 1250$ {\it halos} in the simulations C,D.

Tests performed by Gardini et al. (1999) show that no 
cluster is missed above $4.2 \times 10^{14} h^{-1} M_\odot$.
We have 303 such clusters in A, 316 in B, 40 in C, 36 in D;
6 clusters of C and D where then blown up to various resolution levels.

\section{The virial theorem applied to single clusters: theory}
Finding the volume where cluster materials are in virial equilibrium
may seem a tough and somewhat ambiguous task. If kinetic and potential
energies, within a radius $r$, are defined according to:
$$
2T(<r) = {\sum_i}_{(r_i<r)} mv_i^2 ~,
\eqno (3.1)
$$$$
W(<r) = -{\sum_{i<j}}_{(r_{i,j}<r)} {Gm^2 \over r_{ij}} ~,
\eqno (3.2)
$$
the virial ratio
$$
{\cal R}(<r) = -{2T(<r) \over W(<r)}
\eqno (3.3)
$$
should be unity, in a virialized sphere of radius $r$, provided
that it is fully isolated. On the contrary, in the real world,
as well as in simulations, the virialized volume is bordered by
infalling and outgoing material, possibly traveling through a
partially depleted area. 
For this reason, the convergence of ${\cal R}(<r)$ to unity,
at a precise radius $r$, 
should not even be expected.

In principle, one can take into account such perturbations by
fixing a radius $r$ well inside the cluster volume and 
adding a {\it pressure term} 
to take into account external material,
as has been done in observational
work aiming to estimate the mass of optical clusters
(see, e.g., Girardi et al., 1998). Accordingly, the virial equilibrium
condition reads
$$
2T+W = 3Vp
\eqno (3.4)
$$
($V$ is the volume occupied by virialized material and $p$ is
the pressure at its borders). Eq.~(3.4) translates the problem into finding
a reasonable estimate for $p$.

The problem can also be considered from a slightly different
point of view: if we find, around the cluster, a {\it boundary layer} (BL),
which is in virial equilibrium by itself, and above which virial
equilibrium is lost, its very material and all material it borders,
down to the cluster center, need to be in stationary equilibrium.
Particles may pass through this BL in both directions,
but the overall virial balance shall keep constant in time.

We will now assume that this BL is spherical. 
The conditions found under this restriction are, however, well
approximated by simulated halos.
This seems to indicate that, even after a strongly asymmetric
growth, the final virialization process tends to partially suppress
elliptical features. 

The $r$ dependence of the potential energy can be derived by performing
the sums in eq.~(3.2) starting with the outside:
$$
W(<r) = -{\sum_{i}}_{(r_i<r)} {\sum_j}_{(r_j<r_i)} {Gm^2 \over r_{ij}}  =
$$$$
= -{\sum_{i}}_{(r_i<r)} {\cal Z}(r_i)  ~,
\eqno (3.5)
$$
where the last term is the very definition of $\cal Z$. In the
continuous limit (and still assuming spherical symmetry)
$$
{\cal Z}(r) = Gm \int_0^r d^3r' { \rho(r') \over |r-r'|}
\eqno (3.6)
$$
with a suitable density, $\rho$. Quite in general, a volume integral
of $\rho(r)$ increases with $r$. Henceforth, 
$$
{\cal Z}(r) = C (r/{\bar r})^{-w},
\eqno (3.7)
$$
where C is a normalization constant evaluated 
at an arbitrary position $\bar r$ and $w (< 1)$
will, however, depend on $r$. Accordingly, it must be that
$$
r{d{\cal Z} \over dr} (r)= 
-\big[w + rw'\ln\big({r \over{\bar r}}\big)\big]{\cal Z}(r)  ~.
\eqno (3.8)
$$
Let us now suppose that there exists an interval $\Delta r =r_+ - r_-$,
which satisfies the following conditions:
\begin{itemize}
\item[{(}i{)}]{ it is in virial equilibrium,} 
\item[{(}ii{)}]{ the $r$ dependence of pressure is such that the r.h.s. of
eq.~(3.4), yielding a term of the form
$3(p_+V_+ - p_-V_-)$, can be neglected in the virial balance,}
\item[{(}iii{)}]{ inside $\Delta r$, $w$ is constant. }
\end{itemize}
Further comments on the (ii) requirement can be found in Appendix A,
where we show that halo profiles are expected to approach it.
Together with (i), it implies that
$$
{\sum_i}_{(r_i \in \Delta r)}  mv^2_i -
 r_i {d{\cal Z} \over dr}(r_i) = 0~,
\eqno (3.9)
$$
in the layer.

Let us then take into account the condition 
that $w$ is constant and use eq.~(3.8) with $w'=0$ and the 
{\it virial ratio}
$$
{\cal R}_{\Delta r} = {{\sum_i}_{(r_i \in \Delta r)} mv^2_i
\over {\sum_i}_{(r_i \in \Delta r)} {\cal Z}(r_i) },
\eqno (3.10)
$$
to obtain that
$$
{\cal R} = w ~,
\eqno (3.11)
$$
all along the interval $\Delta r$. 
This equation coincides
with eq.~(3.29) in Goldstein, Pole \& Safko (2002) (in turn,
the latter equation yields eq.~(3.9) hereabove, in the limit
$w \equiv {\rm const.}$).

Requiring that $w'=0$, then, also yields that
$$
{d{\cal R} \over dr} = 0 ~,
\eqno (3.12)
$$
in $\Delta r$. {\it Vice versa}, if the eqs.~(3.11)--(3.12) are 
both fulfilled in a layer of depth $\Delta r$, this layer is at
rest and in virial equilibrium.
We thus define {\it boundary layer (BL)} a region of depth
$\Delta r$ satisfying eqs.~(3.11)--(3.12).

Let us stress that the condition (iii) is essential to set
the $Rw$ requirement, defined by the eqs.~(3.11)--(3.12).
Unless $w$ is constant along a suitable interval, we can
neither replace $r_i d{\cal Z}(r_i)/dr$ by $-w{\cal Z}(r_i)$ 
in eq.~(3.9), nor factorize $w$ out of the sum, so to obtain
eq.~(3.11). Moreover, should eq.~(3.11) hold on a single $r$,
we could have $w'=0$ there, with ${\cal R'} \neq 0$.
Finding a layer fulfilling the $Rw$ requirement, goes therefore
beyond finding a layer in virial equilibrium.
As we shall see, however, once the $Rw$ requirement is fulfilled,
we have found a layer bounding a virialized region.

Let us also stress that the $Rw$ requirement does not imply
that particles cannot
pass through the BL; it only prescribes that such (possible) passages
do not violate its stationary conditions. However, if we find 
about a cluster a boundary layer,
it will act as a confinement barrier to inner kinetic and potential 
energies and mass. Notice that, in principle, inner materials might not 
be in virial equilibrium themselves; even in this case, however, there
can be no net exchange of mass between inside and outside and the
possible energy exchange is limited to the work done by tidal
torques due to anisotropies.

Let us now show that no further material, outside the BL, can be
in virial equilibrium. We will show that this is certainly forbidden
if the values, taken by $\cal R$ and $w$ at $r_+$, are a minimum 
and a maximum, respectively, when compared with values at $r > r_+$.
This simple condition, however, is not strictly necessary; it is
sufficent that the difference ${\cal R}-w$ increases, becoming
positive, at $r>r_+$. This will be shown here below. Let us
however anticipate that, while the
$Rw$ requirement
can be fairly easily tested, it is not easy to deal with such 
further requirement. The tests performed on a large number of clusters
and described in the next sections, however, let us conjecture that,
when the $Rw$ requirement is fulfilled, this further 
requirement is somehow inescapable. 
We now show why, at $r>r_+$, the equilibrium conditions are violated
if a boundary layer is found.
Notice that, at $r > r_+$, we can still assume that $W = W(r_+) 
(r/r_+)^{-w}$, but $w$, starting from the value it had in $\Delta r$,
will now vary with $r$ and, according to eq.~(3.8),
$$
{\cal R } - w = w' r \ln(r/r_+)~;
\eqno (3.13)
$$
this equation cannot be fulfilled, if the difference ${\cal R}-w$,
vanishing up to $r_+$, increases at $r>r_+$, becoming positive.
In fact we required that $w$ is maximum and therefore $w' < 0$.

Let us finally notice that if a layer is characterized by higher $w$
values, this means that it is emptier than inner or outer layers.
A maximum of $w$, therefore, indicates a low density layer. 
In turn, a minimum of $\cal R$ indicates a low kinetic energy, 
with respect to potential energy. BLs, therefore, are partially 
depleted $r$--intervals, where particles, on average, are particularly slow.

\section{The virial theorem applied to single clusters: practical use}
The mathematical definition must now be applied to numerical simulations.
A failure to find BLs could be attributed to insufficient numerical
resolution or to bad violations of spherical symmetry. However, in principle,
even if the resolution is enough and no substantial departures from
sphericity are visible, BLs could still be absent.

Let us now describe our procedure.
After finding clusters in simulations and their centers $C_W$, as 
described in sec.~2, we evaluated $W(r)$ and ${\cal R}(r)$, for spherical
$r$--intervals containing a fixed number 
$$
N = {M_4}/(2^4m_p) ,
\eqno (4.1)
$$
of particles
(here $M_4$ of the total mass within 4.5$\, h^{-1}$Mpc from $C_W$
and $m_p$ is the particle mass).
However, successive points 
along $r$ are obtained by shifting outwards by $N/8$ particles.
In the relevant $r$ range, the above criterion yields $N \sim 10^2$ for 
TCDM (simulation A) and $N \sim 3 \cdot 10^2$ for $\Lambda$CDM (simulation B), 
while successive points lay at a distance $\sim 20\, h^{-1}$kpc. This
is below (but not much below) the force resolution ($\sim 100\, h^{-1}$kpc), 
as desired.

We do not expect BL's to be much thicker than our force resolution;
accordingly, we first seek the minima of $\cal R$. Owing to the 
procedure by which $\cal R$ points are worked out, a minimum is 
considered  {\it significant} only if it is such with respect to 
8 neighbors. Finding the maxima of $w$ is harder, as $w(r)$, obtained
by differentiating $W(r)$, should then be further differentiated to 
find its maxima. In practice, however, this second order differentiation
is not needed. Quite in general, there will be just a few points $r_m$ where 
$\cal R$ has {\it significant} minima. Let then ${\cal R}_m$ be the 
virial ratio at such points and let us fit $W(r)$ with the expression
$$
{\tilde W} = A(r/r_m)^{-{\cal R}_m}
\eqno (4.2)
$$
in $\nu = 8$ points $ > r_m$, covering an interval
$\Delta r  \sim 100$--$150\, h^{-1}$kpc.
The fit is performed in two 
steps: we first minimize the function $\phi 
= W/{\tilde W}-1$ just using $A$ as a parameter; then, in eq.~(4.2), 
we allow ${\cal R}_m$ to take values different from the value of 
$\cal R$ in $r_m$, and perform a 2--parameter fit.

We considered the fit to be satisfactory, if two conditions hold:
\noindent
(i) When performing the first fit, the residual
$$
\chi^2 = \sum_{i=1}^\nu \phi^2(r_i) < 10^{-3}\nu ~,
\eqno (4.2)
$$
(i.e., a mean square discrepancy $\sim 3\, \%$, between the
numerical values of $W(r)$ and its fitting expression,
was our limit of tolerance). 
This detects r--intervals where eqs. (3.11)-(3.12) are both
fulfilled.
\noindent
(ii) When performing the second fit,
the best--fit value of ${\cal R}_m$ is within 1--$\sigma$ from
the actual ${\cal R}_m$ value.
Here we test that also the slope of $W$ is not too far from what is required.
Such constraints should therefore select points where $\cal R$ is minimum and
(nearly) intersects $w$. 

As outlined above,
one might reasonably expect that a large fraction of clusters, at 
their boundaries, is far enough from sphericity that the expected 
system of maxima and minima, as well as their coincidences, is disrupted. 
The numerical noise in calculating $w$ is also quite large
and can be expected to create severe problems. On the contrary, we
found that {\it all clusters with mass $M >\sim 4.2 \cdot 10^{14}h^{-1}
M_\odot$ have at least one value of $r_m$ fulfilling the above 
requirements}, while some clusters have 2 or 3 points where the fitting
criteria are fulfilled. In the few cases, when the fitting criteria were
satisfied at more than one $r_m$, we discarded $r_m$ values yielding
$\Delta_c$ outside an interval $0.3\, \Delta_v$--$3\, \Delta_v$. 
This selection criterion left us with $\sim 97\, \%$ of the initial
clusters. When more than one $r_m$ was still available, we selected
the one corresponding to the smallest $\chi^2$; this last criterion
was needed for $\sim 10 \%$ of the clusters only.

Let us recall again that the requirement of intersection between 
a minimum of $\cal R$ and a maximum of $w$ was found for spherical
systems. When we deal with the halos of our simulations, sphericity
is likely to be an approximation.
When a single intersection exists, however,
it is natural to guess that this is the point where the $Rw$ requirement is
best approached.
However, both in this case and when several $r_m$ points are found,
it is natural to perform a
closer inspection to make sure that no unexpected feature may
bias our understanding of the physical situation.

This closer inspection requires an increased resolution.
As described in sec.~2, a TCDM model and a $\Lambda$CDM model
were then reconsidered within a smaller box with sides of
180$\, h^{-1}$Mpc, making use of the ART and GADGET codes. This gave us
$\sim 1/8$ of the clusters provided by the AP3M simulations,
but enabled us to focus on particular clusters, making use of the
ART package facilities to increase the resolution there while still
keeping identical boundary conditions. In order to test that the
regularities found are not linked to peculiarities of the ART package,
some clusters were magnified using GADGET, instead of ART.
Altogether we have 12 clusters magnified, 6 for each model.
4 of the latter were obtained using ART and 3 using GADGET;
hence, one cluster was magnified with both codes showing that
our results are independent of the n--body code used. 
In fact, the final positions of particles have just minor differences,
while $\cal R$ and $W$ coincide.
A number of subsequent tests also concerned the stability
of the minima of $\cal R$.
In general, the number of local minima $r_m$
decreases when the resolution is increased, but the point
where $W$ fits the expression (4.1) with the worst resolution
is a minimum also with greater resolutions. In just a few cases, new
minima arise when the resolution is improved; but in no case
is a minimum yielding the BL erased when changing resolution.
The ``worst'' cases for each model are shown in Fig.~2. 
Here we have a TCDM cluster, taken from the simulation C,
which passes from 3 minima ($128^3$ particles), to
1 minimum ($256^3$ particles) and, again, to 3 minima
(512$^3$ particles). The BL is displaced outwards
by $\sim 150\, h^{-1}$kpc, when we pass from 128 to 256; then,
when passing from 256 to 512, no appreciable shift occurs.
Out of the 12 clusters considered, only this one presents a shift
slightly above the resolution of the initial simulation.
The  $\Lambda$CDM cluster of simulation D shown in Fig.~2, shows
4 minima with $128^3$ particles, 2 minima with $256^3$ particles and, 
again, to 3 minima with 512$^3$ particles; but when
changing resolution, the shifts are $< 40\, h^{-1}$kpc.

For these 12 clusters we follow in detail the joint
behavior of $\cal R$ and $w$. We show them in Fig.~3 for the
best and worst cases we found. 
Of course, a precise coincidence
between a minimum of $\cal R$ and a maximum of $w$ would
indicate that, within the resolution limits, the system is spherical.
Seeking this kind of coincidence, therefore
is somehow fatuous. Furthermore,
in order to work out $w$, we ought to differentiate $W$,
which is obtained from a discrete point distribution.
Hence, although the nice coincidence of the {\it best} case is 
quite appealing, this case is not {\it safer} than the {\it worst} case
we show. By tolerating a 3$\, \%$ disagreement in $W$, for $r/r_m 
<\sim 1.1$, we allow for a tolerance $\sim 0.3$ on the exponent $w$.
In the worst case, we find a $w$--$\cal R$ discrepancy of this order,
but in most cases, $|w$--$\cal R|$ turns out to be $\sim 0.1$. 
We tentatively conclude that this resolution is sufficient
to show physical features and that 
the level of the $w$--$\cal R$ (dis)agreement 
is a measure of the $a$--sphericity of actual systems.
The important issue is that we {\it always} find a maximum of 
$w$ to match the minimum of $\cal R$, quite close to the $r_m$ selected 
at the initial resolution level.

This point can be further illustrated by Fig.~4, which shows a plot 
similar to Fig.~3, for a typical case at the initial resolution level.
Both the $\cal R$ and the $w$ curves are much noisier here.
$\cal R$, in particular, shows a wide set of local minima; the only
{\it persistent }minima, however, are those marked by thick filled
circles. As expected, the behavior of $w$ is still noisier and most of 
the oscillations shown are clearly numerical. The role of $w$, however, 
just amounts to choosing among the points marked by a thick filled 
circle and selects the point marked by a vertical arrow. 
Even at this resolution level, the presence of a maximum of $w$, 
close to such point, can be suspected. When the resolution 
is increased, this maximum is more clearly visible, while no $w$ maxima
exist close to the other $\cal R$ minima.

A further preliminary test of our technique, concerning galactic-sized and 
unvirialized DM halos, is given in Appendix B.

Before concluding this section, let us draw the reader's attention
on the setting of the BL around particle agglomeration, as shown
in Fig.~5 for a  $\Lambda$CDM cluster. 
This figure shows that the BL is actually set outside of the main
matter agglomeration; its spherical shape is somehow in contrast with
the apparent $a$--sphericity of the cluster materials and easily
justifies minor discrepancies 
from results based
on an assumption of full spherical symmetry.

\section{Results}
Once the sphere confining cluster materials is set, we can 
evaluate the density contrast $\Delta_c$ and the mass $M_c$ inside it.
In Figs.~6 and 7, points give $\Delta_c$ and $M_c$ for all clusters in 
simulations A and B, respectively. They show a rather wide
spread of $\Delta_c$ values. By subdividing
the $M_c$ abscissa in intervals of constant logarithmic width,
we evaluate the average density contrast in each of them,
to seek systematic trends with mass.
Averages are still subject to a significant uncertainty owing to the
spread of $\Delta_c$. They are shown, at the 1--$\sigma$
level, in the plots. At high masses, some logarithmic intervals of
Fig.~7 are empty or contain a single object. In the latter cases 
no error bar is given.

There seems to be no evidence of any peculiar trend of density
contrasts with mass apart from, perhaps, a modest indication of
an increasing density contrast at very high scales in $\Lambda$CDM.
It is therefore possible to consider the overall average among
$\Delta_c$'s.  This average is indicated by the continuous horizontal
line and compared with the "virial" density contrast $\Delta_v$, as 
given by eq.~(1.1). In both cases, $\Delta_v$ (dotted line) is well 
inside the range of the density contrasts we found; the average
$\Delta_c$, however, in both cases, is smaller than $\Delta_v$
by $\sim 25\, \%$.

It is, however, clear that gravitationally
evolved, stationary objects are {\it not} uniquely identified with
a given, fixed density contrast. 
In further work, we plan to deal with the impact of variable
$\Delta_c$ on the physical characteristics of DM halos, such
as concentration, average angular momentum, density
and velocity profiles.  It may also be significant
to test our results against the effects of different particle 
selection criteria, e.g. using the SKID algorithm instead of the SO
one to pre-select the DM halos.

In Figs.~8 and 9 we plot $r_c$ against $M_c$, for the simulations
A and B, respectively. Here, error bars account for the variance 
of $r_c$ around its average, which is wider than
the error of the average $r_c$, shown in Fig.~6. 
Continuous and dotted curves show the expected behavior
of $r_c$ against $M_c$, when the density contrast is set either
at the average $\Delta_c$ value or at $\Delta_v$.

It is also significant to compare the plot of $r_c$ against $M_c$
with the behaviour of the radii $r_c$ against the
masses $M_v$ of the corresponding clusters. This comparison is shown
in Fig.~10, for the simulation A only; in order to avoid confusion, we 
omit error bars which, however, are approximately of the same size 
in both cases. The highest mass points, for which error bars cannot
be evaluated, are also omitted. Filled circles yield $r_c$ $vs$. $M_c$,
empty circles yield $r_c$ $vs$. $M_v$. Filled circles nicely fit the
line worked out from the average density contrast $\Delta_c$; only
at high masses, where the statistics is poorer, some of the circles
are far from this line. The same line is also a reasonable fit for
the open circles. These, however, are systematically farther
at all mass scales. In the figure, however, we also plot an horizontal
line showing the average of the $r_c$ values of all clusters with
mass $>4.2 \cdot 10^{14} h^{-1} M_\odot$. Even this average can  be
considered a reasonable fit for open circles (an unweighted $\chi^2$ 
evaluation gives similar probabilities to both fitting lines).
This plot shows that there are actual dangers, if quantities
worked out assuming a constant $\Delta_v$ are compared with real
data. In this case, one might tentatively suggest that there is a 
typical, mass independent, cluster radius $R \simeq 1.56\, h^{-1}$Mpc,
quite close to the Abell value. Our procedure, instead, outlines that
this arises because some clusters were attributed a biased
mass and the corresponding points were then set at a wrong abscissa.
Although, altogether, this abscissa reshuffling is substantially
casual, the danger that an characteristic scale erroneously 
appears is far from absent.

A critical result of our analysis is however shown in Figs.~11 and 12
(for simulations A and B, respectively).
In simulations, the ``numerical" cluster masses can be easily evaluated
by summing up particle masses. The actual approach, in the real
world, ought to be quite different and is based, first of all, on
an analysis of galaxy velocities or X-ray emitting gas temperature.
These analysis make use of the virial hypothesis; therefore, measured
masses of galaxy clusters are estimates of their {\it dynamical} masses.
It is therefore important to test how ``numerical" cluster masses 
in simulations agree with the masses 
$$
M_{dyn}=  {{ \langle v^2 \rangle_{\{v,c\}} \cdot r_{\{v,c\}}}  \over G}~,
\eqno (5.1)
$$
evaluated averaging over the velocities of the particles within $r_{v,c}$.
This comparison has been often performed, showing
a reasonable coincidence between $M_v$ and $M_{dyn}$. The results
of such fit are shown in Figs.~11 and ~12 by the dashed histograms, which
confirm the slight excess of $M_{dyn}$ $vs$. $M_v$ already noticed
by previous authors. In the same plots we also show the results
of a fit between $M_{dyn}$ and $M_c$, obtained on the basis of the
setting of the BL. There seems to be little doubt that the
latter fit is better. The average values of $M_{dyn}/M_c$ are both
$\sim 0.97 \pm 0.03$, quite close to unity
and consistent with it at the 1--$\sigma$ level. 

We have also compared the shapes of the cluster (integral) mass functions 
$n_c(>M_c)$ and $n_c(>M_v)$. In Fig.~13 and 14, upper and lower plots
show their behaviors, respectively. In the same plots, we also show 
Press \& Schechter and Sheth \& Tormen theoretical curves. They can be
obtained from two different expressions of
$$
f(\nu)\nu \, d\log \nu = {M \over \bar \rho_m} n_c(M) M\, d\log M~;
\eqno (5.2)
$$
here $\bar \rho_m$ is matter density, the bias factor $\nu = \delta_c/
\sigma_M$ is the ratio of the critical overdensity in the spherical 
growth model to the rms density fluctuation on the length scale 
corresponding to the mass $M$, and $n_c(M)$ is the differential 
mass function, which shall then be integrated to obtain $n_c(>M)$.

In the Press \& Schechter formulation, assuming spherical unperturbed
fluctuation growth,
$$
f(\nu)\, \nu = \sqrt{2/\pi}\, \,  \nu \exp(-\nu^2/2) ~,
\eqno (5.3)
$$
while Sheth \& Tormen, taking into account the ellipsoidal nature
of the collapse, argue that
$$
f(\nu) \, \nu = A(1+\nu'\, ^{-2q}) \sqrt{2/\pi}\, \,  
\nu' \exp(-\nu'\, ^2/2) ~,
\eqno (5.4)
$$
with $\nu'=\sqrt{a}\, \nu$; this expression introduces the parameters 
$A$, $q$ and $a$, however constrained by the requirement that
all matter is bound in objects of some mass $M$, however small,
and therefore the integral of $f(\nu)$ must be unity. In our plots,
$A$ is normalized to yield the number of clusters found in simulations, and
this constraint is automatically satisfied. Sheth \& Tormen (1999)
showed that, in the GIF simulations for an SCDM, 0CDM and $\Lambda$CDM 
models (see Kauffmann et al 1999), a good fit to the numerical
mass function was obtained if $q = 0.3$, $a=0.707$. 

In Fig.~13 and 14 we report the mass function obtained from eq.~(5.4) for
these values, for TCDM and $\Lambda$CDM, respectively. The fit 
with the $n_c(>M_v)$ is surely better than using
eq.~(5.3). We also tried to modify the parameter $a$, that, according
to Sheth \& Tormen arguments, is related to the density contrast
of virialized structures. As a matter of fact  $n_c(>M_c)$ is derived
with a variable density contrast, as explained in previous sections,
and, however, with an average final density contrast which is
$\sim 25\, \%$ below $\Delta_v$. It is therefore reasonable to 
expect that a lower value of $a$ is needed to compare
theoretical predictions with the numerical behavior of $n_c(>M_c)$.
In Fig.~13 and 14, we show the expression (5.4), both for 
$a = 0.707$ and 0.5. It is also evident that, while $a=0.707$ provides
a good fit for $M_v$--mass functions, the fit obtained provided by 
$a=0.5$ for the $M_c$--mass function is perhaps better.

The results of these comparisons, altogether, are that:
(i) No major variations in the numerical mass function
arises when $M_v$ is replaced by $M_c$. (ii) Variations
however exist, which are of the same order of the differences
between the standard theoretical expression (5.3; Press \& Schechter) 
and the improved theoretical expression (5.4; Sheth \& Tormen). 
(iii) When the latter expression is used, the fits to the numerical 
mass functions are best if different values of the parameter 
$a$ are used for $M_v$ and $M_c$ ($\sim$0.7 and 0.5, respectively).
The differences between the mass functions $n_c(>M_v)$ 
and  $n_c(>M_c)$ are more evident in the intermediate mass range 
($7 \cdot 10^{14} h^{-1} M_{\odot} < M < 10^{15} h^{-1} M_{\odot}$).
They arise from a widespread ``re-shuffling" among the masses 
of DM halos, when the two definitions are used, causing thereby 
different numbers of halos in different mass intervals.
This can also be seen from Fig.~10,
where the mean cluster radii ($r_c$, $r_v$) of halos in this mass range show
appreciable differences.

\section{Conclusions}

Observational work on galaxy clusters made a giant leap forward
thanks to X--ray observations and, in turn, cluster data are
providing more and more cosmological information. The ROSAT All
Sky Survey (RASS) allowed the extraction of large samples of galaxy clusters,
like REFLEX (Boehringer et al. 2001) or NORAS (Viana et al. 2001) , 
tracing the large scale structure (LSS) of
the Universe up to $\sim 1000\, h^{-1}$Mpc of distance ($h$ is
the Hubble constant in units of 100 km/s/Mpc). These data are 
consistent with the spectrum of a low--density CDM model, with
matter density parameter $\Omega_m \simeq 0.3$. Also an analysis
of the redshift dependence of the X--ray cluster luminosity
function, based on a Press \& Schechter approximation, sets a
relation between $\Omega_m$ and $\sigma_8$, which requires again $\Omega_m
< 0.5$, 0.6 ($\Omega_m$: matter density parameter; $\sigma_8$: squared
mean density contrast, on the scale of 8$\, h^{-1}$Mpc) (Borgani et
al. 2001). Such constraints
on the model will become more stringent, in the forthcoming years, thanks
to XMM--Newton and Chandra data, providing high--$z$ cluster samples
with tenths of thousands objects (see, e.g., Bartlett et al. 2001).

In our opinion, such fast improvements of the observational
materials deserve to be matched by a suitable effort of the 
theoretical analysis. In this work we try to contribute to it,
by providing an improved
pattern to measure individual cluster densities, radii and masses
in numerical simulations, without reference to a standard 
density contrast. 
Although, both in real samples as well as in simulations, cluster
members are directly observable together with their environs,
a definition of cluster radii, masses, and density contrasts
is needed both for statistical aims, when treating a large set of
halos, and for individual objects, e.g., to compare the observed
mass distribution with analytical expressions of radial profiles.
In the standard approach, a value for the density contrast is
usually borrowed from analytical results,
which are obtained under assumptions which can be incorrect 
in the real world, as well as in simulations.
Such assumptions, however, were explicitly considered to be the
best possible choice.
In particular, Cole \& Lacey (1996) had explored the possibility
to make use of the virial ratio, but considered its value in
spheres, instead of layers. Here we see that, passing
from the integral to the differential virial ratio 
${\cal R} (r) = -2T(r)/W(r)$, allows the detection of a neat
boundary, where a transition from halo to surrounding material 
occurs. In fact, first of all, (i) this inspection allows the
detection of
minimum points of $\cal R$ in all halos, which could be easily missed 
in an integral inspection.
Some of them are safely shown not to be numerical features, as they
persist without appreciable shifts  when we compare different
resolution simulations of a given halo. Furthermore, (ii) we found 
that, 
CHb 
quite close to the radius $\bar r$, where one of the persistent minima
occurs, there is 
always
a maximum of the function $w(r)$; finally, 
CHb
(iii) at the radius $\bar r$, we have also that
$$
w({\bar r}) \simeq {\cal R}({\bar r})~.
\eqno (6.1)
$$ 
Let us recall that, in general, $w(r)$ is defined with reference 
to the behavior of the potential energy around $\bar r$, by setting:
$$
W(r) = W({\bar r}) (r/{\bar r})^{-w(r)} 
\eqno(6.2)
$$
and, therefore, its $r$ dependence accounts for the radial
dependence of $W(r)$.

In this paper we have shown that the above properties (i), (ii), 
(iii) are consistent with the requirements that:
(a) $\bar r$ sets a neat boundary between bound and unbound material.
(b) A significant interval (depth $\sim 50$--100$\, h^{-1}$kpc), 
where $w(r)$ is constant, lies about $\bar r$. We called
it a boundary layer (BL). Accordingly, the $r$ dependence of
$W(r)$, inside of the BL, is expressed by eq.~(6.1),
by replacing the exponent $w(r)$ with its constant value
$\bar w$.

The relation between the observed features (i), (ii), (iii) and the 
properties (a), (b), was shown under the assumption of spherical symmetry. 
The fact that the conditions (i), (ii), (iii) are not satisfied exactly,
but with a (slight) approximation, can be easily
attributed to deviations of numerical structures from sphericity.
However, such deviations ought to be quite mild, owing to the
statistics we performed on a large number of halos,
and this indicates that
violent relaxation erased most previous $a$--spherical features.
Non--spherical collapses, however, 
leave a substantial imprint on the final density contrasts 
$\Delta_c$ found for virialized halos 
whose spread accounts for different initial 3--D geometries
of fluctuations and different interactions with the environment
during the collapsing stages.

The simulations used to focus the above properties, at different
levels of resolution, in TCDM and $\Lambda$CDM modes, were run 
both with the GADGET and the ART codes. Results do not depend on
the code. We could also extend the analysis, at the lowest
resolution level, to simulations run in quite large boxes
(360$\, h^{-1}$ Mpc aside), where we had a large statistics of
clusters ($> 300$ with $M_c > 4.2 \cdot 10^{14} h^{-1}M_\odot$,
for each model). We applied to all of them an $Rw$--requirement, 
meant to detect the BL, and found it in 97$\, \%$ of the clusters
considered. {\it A detailed inspection of these clusters shows that 
the mass $M_c$ within the BL substantially coincides with the
mass $M_{dyn}$, evaluated from the velocities of all particles within
$r_c = \bar r$, according to the virial theorem.} Furthermore,
a direct inspection of the clusters without a BL,
showed they were exceptional structures, mostly due to undergoing
major merging, which certainly justified a dynamically unsettled
condition and a badly $a$--spherical geometry. Let us however
notice (see Fig.~5) that ``standard'' $a$--sphericity of
halo structures does not appear to be a serious obstacle
to BL detection.

At present, model clusters to fit X--ray data are mostly built
using hydro--codes. This allows substantial improvements with respect to
initial analyses based on pure Newtonian dynamics. There are, however,
some features in the interpretation of both kinds of simulations,
which rely on a sound definition of the virialization radius $R_c$.
This radius, whose setting immediately follows from the
detection of the BL position,
corresponds to a wide spread of density contrasts $\Delta_c$,
and, in general, is significantly different from the
radius $R_v$, obtained using a given density contrast $\Delta_v$.
Even keeping to models without hydrodynamics, it is clear that
using different radii $R_c$ and density contrast $\Delta_c$ 
could lead to different relations between $L_X$, $T_X$ and $M_c$. 
As a consequence the one--to--one correspondence between $M_c$ and 
$T_X$ could be lost and it would no longer be granted that a 
mass function and
a temperature function are substantially equivalent quantities.

\section*{ACKNOWLEDGEMENTS}

We thank INAF for allowing us the CPU time 
to perform the ART simulation C and D at the CINECA consortium 
(grant cnami44a on the SGI Origin 3800 machine).
GADGET simulations of high--resolution clusters have
been run on the 16 Linux PC Beowulf cluster at the Osservatorio
Astronomico di Torino.
Alessandro Gardini is thanked for allowing us to make
use of the simulations A and B; we also
wish to thank him, Loris Colombo and Giuseppe Tormen for useful
discussions during the preparation of this paper.

\vfill\eject

\section*{APPENDIX A}

Let us assume that halos are spherically symmetric. If BL's are 
in virial equilibrium, at any $r$, in their interior
$$
\langle v^2 \rangle = -\phi = G {M(<r) \over r},
\eqno (A1)
$$
$M(<r)$ being the mass inside $r$. According to kinetic gas theory, 
however, $p=\rho \langle v^2 \rangle/3$. Accordingly, local equilibrium 
prescribes that $p=-\rho \phi/3$ and, up to first order in $\Delta r/r$,
$$
3[p_+V_+ - p_-V_-]
= {4\pi \over 3} G \rho_- r_-^{3-\beta} \big[  {M_+ \over r_+^{1-\beta}}
- {M_- \over r_-^{1-\beta}} \big] \simeq ~~~~~~
$$
$$
~~~~~~~~~~~~~~~\simeq 4\pi G \rho_- r_-^3 {2\beta-1 \over 3}
{M_- \over r_- }{\Delta r \over r_-}
\eqno (A2)
$$
with a $\beta$ value such that $\rho r^{3-\beta}$ is constant
across $\Delta r$ (indices $_\pm$ indicate that a quantity
is evaluated in $r_\pm$).

Let us compare this term with the potential energy of the layer
$$
W = G \int_{r_-}^{r_+} d^3r\, \rho(r) {M(<r) \over r}
\simeq 4\pi G \rho_- r_-^3 {M_- \over r_- }{\Delta r \over r_-},
\eqno (A3)
$$
also evaluated assuming spherical symmetry and
neglecting terms of higher order in $\Delta r/r$.
The ratio between $pV$ terms and potential energy is 
$\simeq (2\beta-1)/3$ and would vanish for $\beta = 0.5$.

Let us now assume to be dealing with halos whose profile is 
$\rho(r) \simeq \rho_o \chi (r/r_s)$ with
$$
\chi(x) = [x^\alpha(1+x)^{3-\alpha}]^{-1} ~,
\eqno (A4)
$$
where $r_s$ is the scale radius of the profile and the
concentration of the halo is $c=r_s/r_v$.
This profile is found for most halos in our simulations, up to 
$r < r_v$, with a value of $c$ in the interval 
4--7 and values of $\alpha \simeq 1$; this agrees with previous
numerical analyses (see, e.g., Navarro et al. 1996, 1997, Ghigna et al 
2000, Klypin et al 2001, Power at al 2002), also extended 
to smaller values of $r$,
finding values of $\alpha$ ranging between 0.8 and 1.2.
Using this profile we can evaluate the $r$ dependence of $\beta$
for $\alpha$ values in the above interval. Such values are shown in
Fig.~15 and indicate that for $c \sim 4$--7, $\beta \sim 0.5$--0.3.

This shows that the contribution of the pressure terms
to the virial balance is expected to be significantly smaller than the
contribution of the potential term.

Let us however add some further comments:

(i) Using the profile in eq.~(A4) for $r >\sim r_v$ is an approximation.
An inspection of numerical halos, already significantly inhomogeneous
at such radii (see, e.g., Fig.~5), shows that the slope of the profile 
is often smoother there.

(ii) Accordingly, owing to eqs.~(A2)--(A3), 
the $pV$ terms can be absorbed in a factor ($\sim 1$), 
set in front of the potential term of the virial balance. 
This correction does not cause any displacement of the point
where $\cal R$ has its minima. On the contrary, it may interfere
with the condition $w = \cal R$, which, in principle, can be
suitably improved.

(iii) Deviations from spherical symmetry, however, are far from
being fully negligible and the discrepancies between $\cal R$
and $w$, found in halo analysis, show that a correction of $\cal R$
by $\sim 10\, \%$, would not improve the efficiency of the method.

\section*{APPENDIX B}

We tried to find a BL, i.e. to apply the $Rw$ requirement, also in 
two N--body simulations aimed at studying the evolution of a galactic 
stellar disk in a cosmological context. These simulations, performed 
with GADGET, will be discussed in a forthcoming paper (Murante, Curir 
\& Mazzei, in preparation). Here we used simulations with DM only (no 
galactic disk is present). 

In the first one, initial conditions were set up with the same 
multi--mass technique described in Sec.~2, using the ART package.
A DM halo, whose mass is $M_v=1.14 \times 10^{11} h^{-1} M_\odot$ 
at redshift $z=0$, was simulated at high--resolution using DM particles 
of mass $M_p=1.21 \times 10^6 h^{-1} M_\odot$. According to eq.~(1.1), 
the virial radius $r_v \approx 0.125 h^{-1}$ Mpc. External forces were 
taken into account by setting heavier and heavier particles in 3 concentric 
shells. We checked that no intruders (heavier mass particles) were ever 
closer than 0.5 $h^{-1}$Mpc from the center of mass of the halo. 
The $Rw$ requirement was applied to this halo, at redshifts $z=0$ 
and $z=2$.

At $z=0$ a BL was found, with a radius $r_c=0.113 h^{-1}$ Mpc, yielding
a ratio $M_c/M_{dyn}=1.0502$. On the contrary, at z=2, we could not find 
a good fit for the $\chi^2$ parameter used to implement the $Rw$ requirement.
It should be noticed that these results held, in spite of the fact that
the halo concentration is $\simeq 20$.

This suggests that BL's are found by the $Rw$ requirement also when
the condition $pr^3 = {\rm const.}$ is (mildly) violated; on the
contrary, no BL is found when the {\it the halo is not yet virialized}, 
as it was at z=2 and is also confirmed by direct inspection. Apparently, 
the presence of a BL however distinguishes gravitationally relaxed
objects from those still in an evolutionary phase.

A partial confirm of these two results comes from the application of the
$Rw$ requirement to a second simulation, following the evolution of an 
isolated halo of nominal mass equal to the previous one, with a nominal 
radius of $R=0.180\, $Mpc, generated with a NFW density profile
(see eq.~A4) and with initial particle velocities picked up from a
multivariate Maxwellian distribution; the NFW density profile 
was cut off at the nominal radius; the concentration $c$ was set 
to the value measured in the previous case (using $r_v$ to evaluate 
it, as in Navarro et al, 1997). The halo was evolved for $\approx 10\, 
$Gyr, to reach the age of the Universe at $z=2$ in a $\Lambda$CDM 
cosmology where $\Omega_\Lambda=0.7$ and $\Omega_m=0.3$. 

We applied the $Rw$ algorithm, to find a BL at the beginning ($t=0\, $Gyr) 
and at the end ($t=10.23\, $Gyr) of the simulation. No BL could be 
found at $t=0$, when the halo was {\it not} in equilibrium. Apparently,
therefore, the presence of a BL is characteristic of gravitationally 
stable objects, independently of the geometrical distribution of particles.
At $t=0$, a SO algorithm, based on a purely geometrical prescription,
would comfortably detect a halo. The $Rw$ requirement, instead, which is
based on dynamical prescriptions, appears to be more selective than purely 
geometrical recipes.

At $t=10.23$ Gyr, we find a layer fulfilling the $Rw$ requirement, with
$r_c=0.023\, $Mpc, yielding $M_c/M_{dyn} = 1.0201$. This happens in spite
of the value of the density contrast at $r_c$, which is quite large
(${\rho/ \rho_{crit}}=2457$). In this (quite peculiar) case, the 
definition of virialization based on the density contrast fails,
in spite of the fact that gravitational equilibrium is established.

These arguments seem to indicate that a technique based on the $Rw$ 
requirement is applicable on a wide range of DM halo masses, even
when $c$ is fairly large; mild deviations from the requirement that
$pr^3$ is constant do not seem to destroy its effectiveness. Further
analysis will be performed to put precise limits to the cases when 
the $Rw$ criterion is a useful tool for telling apart virialized halos from
still strongly evolving ones. 

\clearpage

\begin{table}
\begin{center}
\begin{tabular}{l|c|c}
\multicolumn{3}{c}{} \\
&\multicolumn{1}{c}{TCDM (A)}&\multicolumn{1}{c}{$\Lambda$CDM (B)}
\\
\hline \\
 $\Omega_m$ &      1   & 0.35\\
 $\Omega_{\Lambda}$ &       0   & 0.65\\
 $\Omega_b \cdot 10^2$ &      6   & 3.6\\
 $n$                 &     0.8  & 1.05\\
 $h$ &       0.5   & 0.65\\ 
 $\sigma_8$            &    0.55  & 1.08 \\
 $\Gamma$             &   0.32  & 0.19  \\
\multicolumn{3}{c}{} \\
\multicolumn{3}{c}{} \\
\end{tabular}
\caption{Parameters for the simulations performed by the AP3M code.}
\label{tab1}
\end{center}
\end{table}

\begin{table}
\begin{center}
\begin{tabular}{l|c|c}
\multicolumn{3}{c}{} \\
&\multicolumn{1}{c}{TCDM (C)}&\multicolumn{1}{c}{$\Lambda$CDM (D)}
\\
\hline \\
 $\Omega_m$ &      1   & 0.3\\
 $\Omega_{\Lambda}$ &       0   & 0.7\\
 $\Omega_b \cdot 10^2$ &      5   & 2.6\\
 $n$                 &     0.8  & 1.05\\
 $h$ &       0.5   & 0.7\\ 
 $\sigma_8$            &    0.55  & 1.08 \\
 $\Gamma$             &   0.31  & 0.21  \\
\multicolumn{3}{c}{} \\
\multicolumn{3}{c}{} \\
\end{tabular}
\caption{Cosmological Parameters for the ART simulations.}
\label{tab2}
\end{center}
\end{table}

\clearpage

\begin{figure}
\centerline{\mbox{\epsfysize=7.0truecm\epsffile{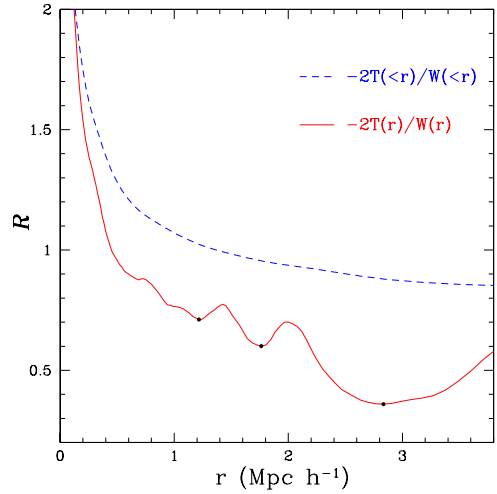}}}
\caption{Integral (dashed line) and differential (solid line) 
virial ratios ${\cal R}(>r)$
and  ${\cal R}(r)$ are plotted against the halo radius $r$
for a $\Lambda$CDM halo. Heavy dots indicate the points where
the differential $\cal R$ has minima.}
\label{Fig 1}
\end{figure}

\begin{figure}
\centerline{\mbox{\epsfysize=7.0truecm\epsffile{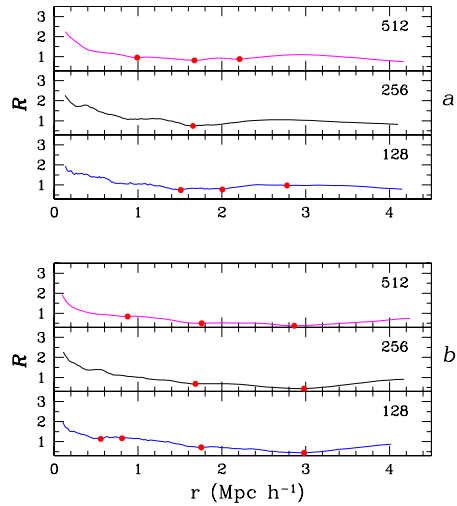}}}
\caption{The virial ratio ${\cal R}=-2T/W$ is plotted against
the halo radius $r$, at different resolution levels. The value
of $N$ shown in each box yields the number $N^3$ of particles
used at the highest resolution level. Heavy dots show the
setting of the {\it permanent} minima. The plots (a) and (b)
refer to $\Lambda$CDM and TCDM clusters, respectively. The minima
fulfilling the $Rw$ requirement lay at $\sim 1.62\, h^{-1}$Mpc
and at  $\sim 1.75\, h^{-1}$Mpc for the halos $a$ and $b$,
respectively.}
\label{Fig 2}
\end{figure}

\begin{figure}
\centerline{\mbox{\epsfysize=7.0truecm\epsffile{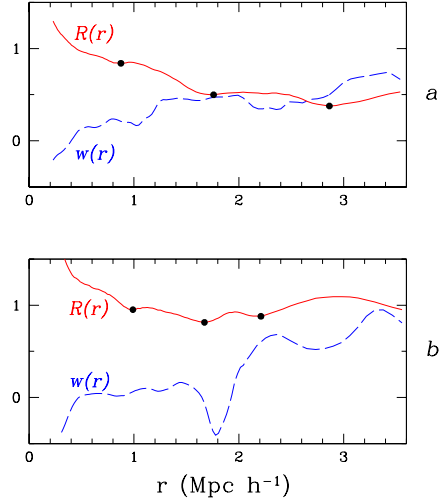}}}
\caption{The $\cal R$ and $w$ dependence on $r$ is shown in the
best and worst cases treated at the highest resolution level, ($a$ and $b$,
respectively); still at this resolution, the $w$ behavior is
rather noisy (compare however with Fig.~4, to see how better 
resolution reduces numerical noise). According to the $Rw$
requirement, the case $a$ shows a neat intersection of the
two curves for a maximum of $w$ and a minimum of $\cal R$.
However, also in the case $b$, the correspondence between the
maximum of $w$ and the minimum of $\cal R$ is apparent.
}
\label{Fig 3}
\end{figure}

\begin{figure}
\centerline{\mbox{\epsfysize=7.0truecm\epsffile{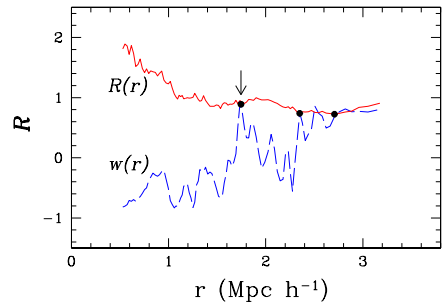}}}
\caption{A typical example of $\cal R$ and $w$ behaviors at the
initial resolution level. This halo is the same shown in Fig.~3$a$,
at higher resolution.
}
\label{Fig 4}
\end{figure}

\begin{figure}
\centerline{\mbox{\epsfysize=7.0truecm\epsffile{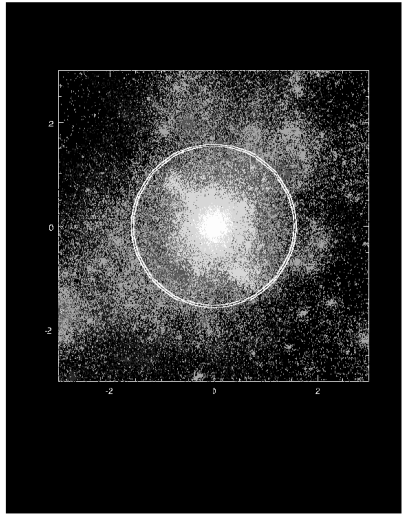}}}
\caption{Image of a typical cluster, with the location of the boundary 
layer (BL). The cluster shown has a radius $r_c = 1.76 \, h^{-1}$Mpc, 
a mass $M_c = 8.92 \cdot 10^{14} h^{-1} M_\odot$ and a density
contrast $\Delta_c = 141$. The m.s. velocity of particles within $r_c$
is 1574$\, $km/s. This cluster is taken from simulation D. 
Gray scales refer to particle velocities.
}
\label{Fig 5}
\end{figure}

\begin{figure}
\centerline{\mbox{\epsfysize=7.0truecm\epsffile{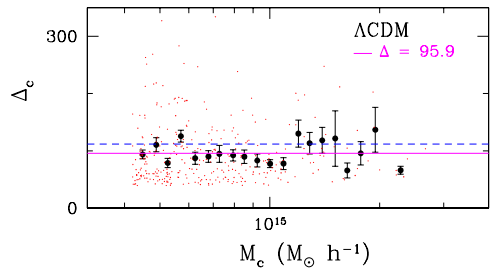}}}
\caption{The mass dependence of the density contrast is shown.
Points refer to all clusters in the simulation B ($\Lambda$CDM).
Heavy points give the average $\Delta_c$ for each $M_c$ interval
(constant logarithmic width). Bars are 1--$\sigma$ errors of the
averages. The thick horizontal line is the average value of
$\Delta_c$. The dashed line is the virial density contrast
$\Delta_v$ expected for a unperturbed spherical fluctuation
growth.
}
\label{Fig 6}
\end{figure}

\begin{figure}
\centerline{\mbox{\epsfysize=7.0truecm\epsffile{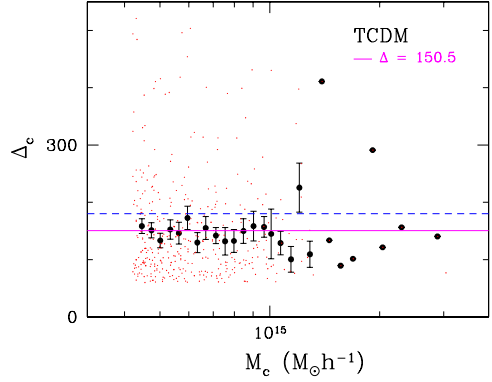}}}
\caption{The same as Fig.~6 for clusters in simulation A (TCDM).
At high $M_c$ some heavy points are shown without error bars,
when there is one cluster per logarithmic mass interval.}
\label{Fig 7}
\end{figure}

\begin{figure}
\centerline{\mbox{\epsfysize=7.0truecm\epsffile{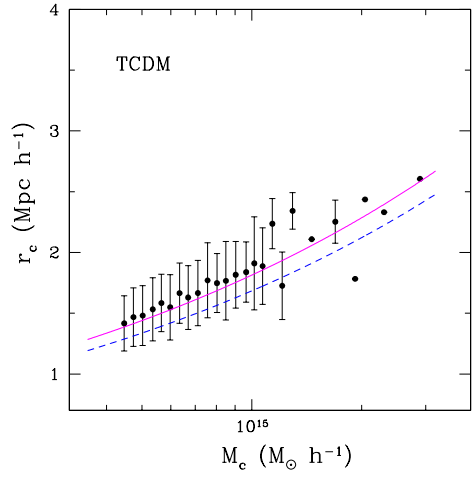}}}
\caption{Average cluster radii (heavy dots) are plotted against
their masses $M_c$ for the simulation A. The solid line is obtained
for the average density contrast (150.5); the dashed line, instead, is
obtained from the ``virial'' density contrast (178). Error
bars yield the variance of cluster radii around their averages. }
\label{Fig 8}
\end{figure}

\begin{figure}
\centerline{\mbox{\epsfysize=7.0truecm\epsffile{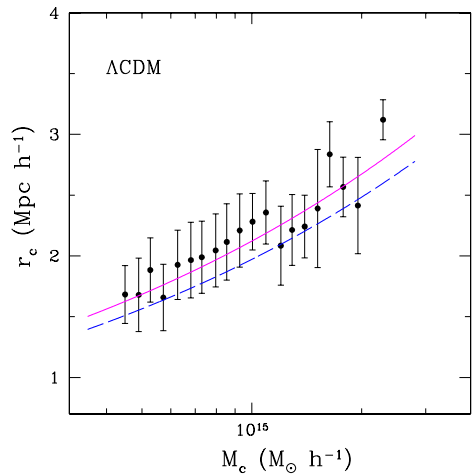}}}
\caption{The same as Fig.~8, for the simulation B. Solid and dashed
lines correspond to density contrasts 95.9 and 111, respectively.
}
\label{Fig 9}
\end{figure}

\begin{figure}
\centerline{\mbox{\epsfysize=7.0truecm\epsffile{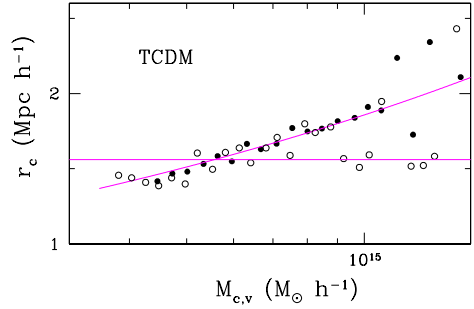}}}
\caption{We show how ``virial'' quantities can introduce spurious 
characteristic scales, if fit against
``real'' quantities. Here mean cluster radii ($r_c$) are plotted
against $M_c$ (filled circles) and $M_v$ (empty circles).
Error bars are omitted to avoid confusion (see, however, Fig.~7).
The horizontal line could be interpreted as a mass--independent 
average radius for all clusters.
}
\label{Fig 10}
\end{figure}

\begin{figure}
\centerline{\mbox{\epsfysize=7.0truecm\epsffile{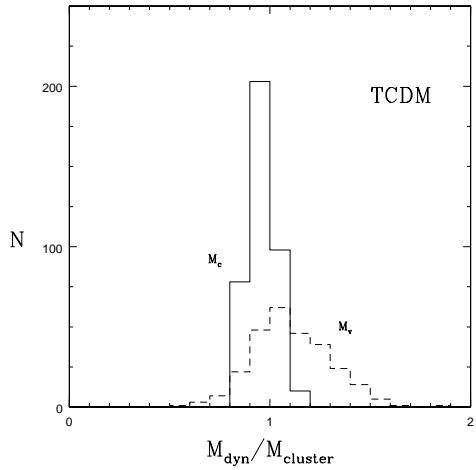}}}
\caption{The ratio between masses evaluated from particle velocities,
according to the virial theorem ($M_{dyn}$), and summing particle
masses ($M_{cluster}$) is shown limiting clusters either at $r_c$ or
at $r_v$, for all clusters of simulation A.
}
\label{Fig 11}
\end{figure}

\begin{figure}
\centerline{\mbox{\epsfysize=7.0truecm\epsffile{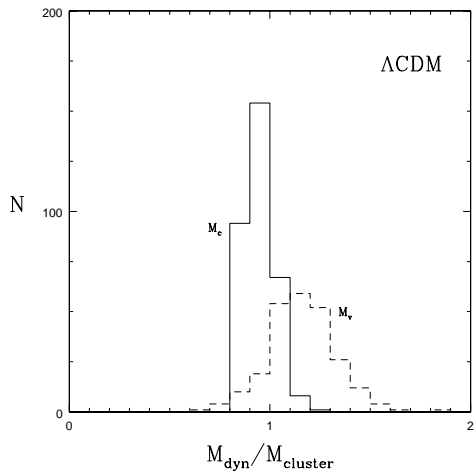}}}
\caption{The same as Fig.~11 for simulation B.
}
\label{Fig 12}
\end{figure}

\begin{figure}
\centerline{\mbox{\epsfysize=7.0truecm\epsffile{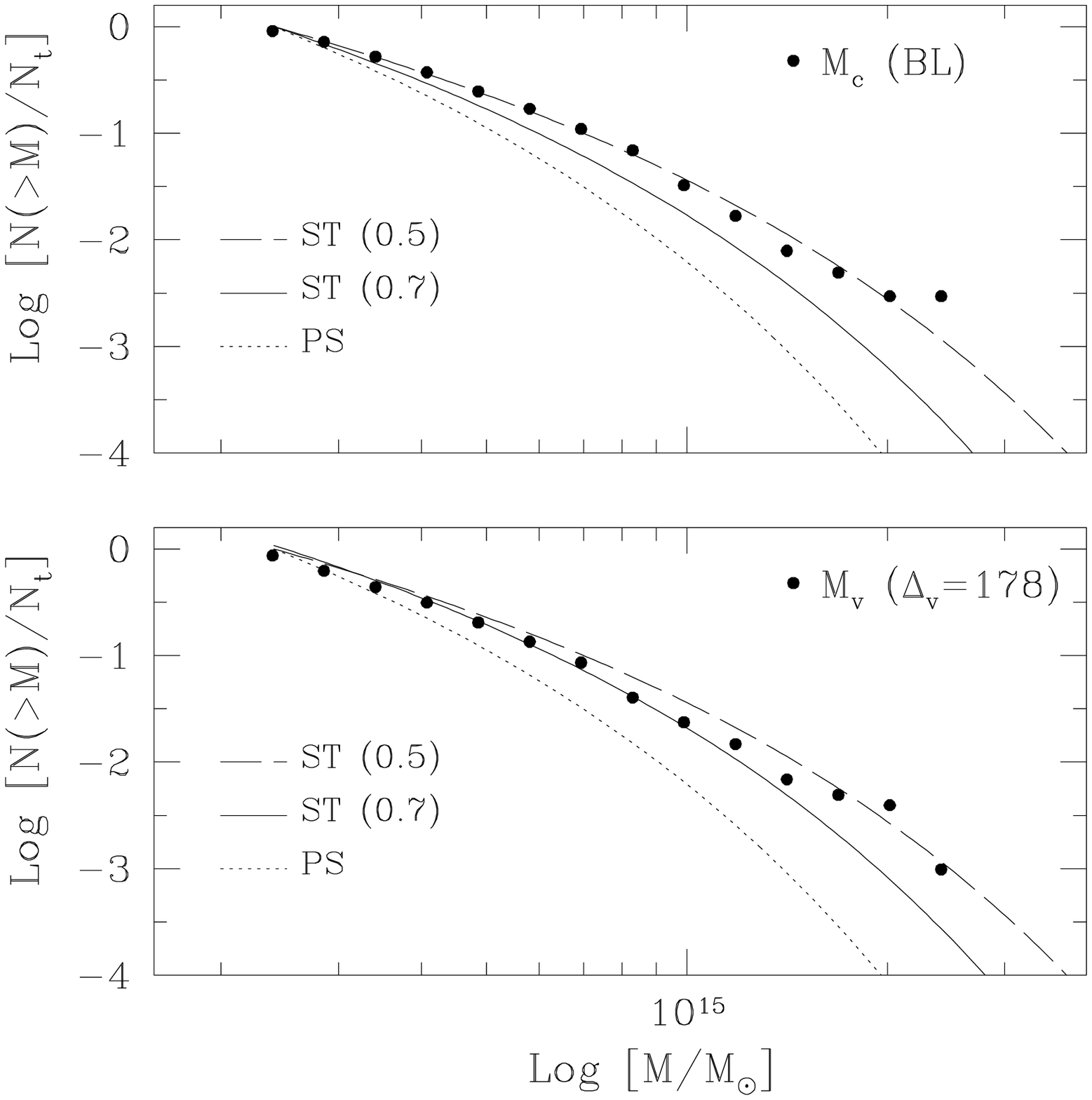}}}
\caption{Comparison between mass functions obtained using BL
or virial quantities (TCDM).
}
\label{Fig 13}
\end{figure}

\begin{figure}
\centerline{\mbox{\epsfysize=7.0truecm\epsffile{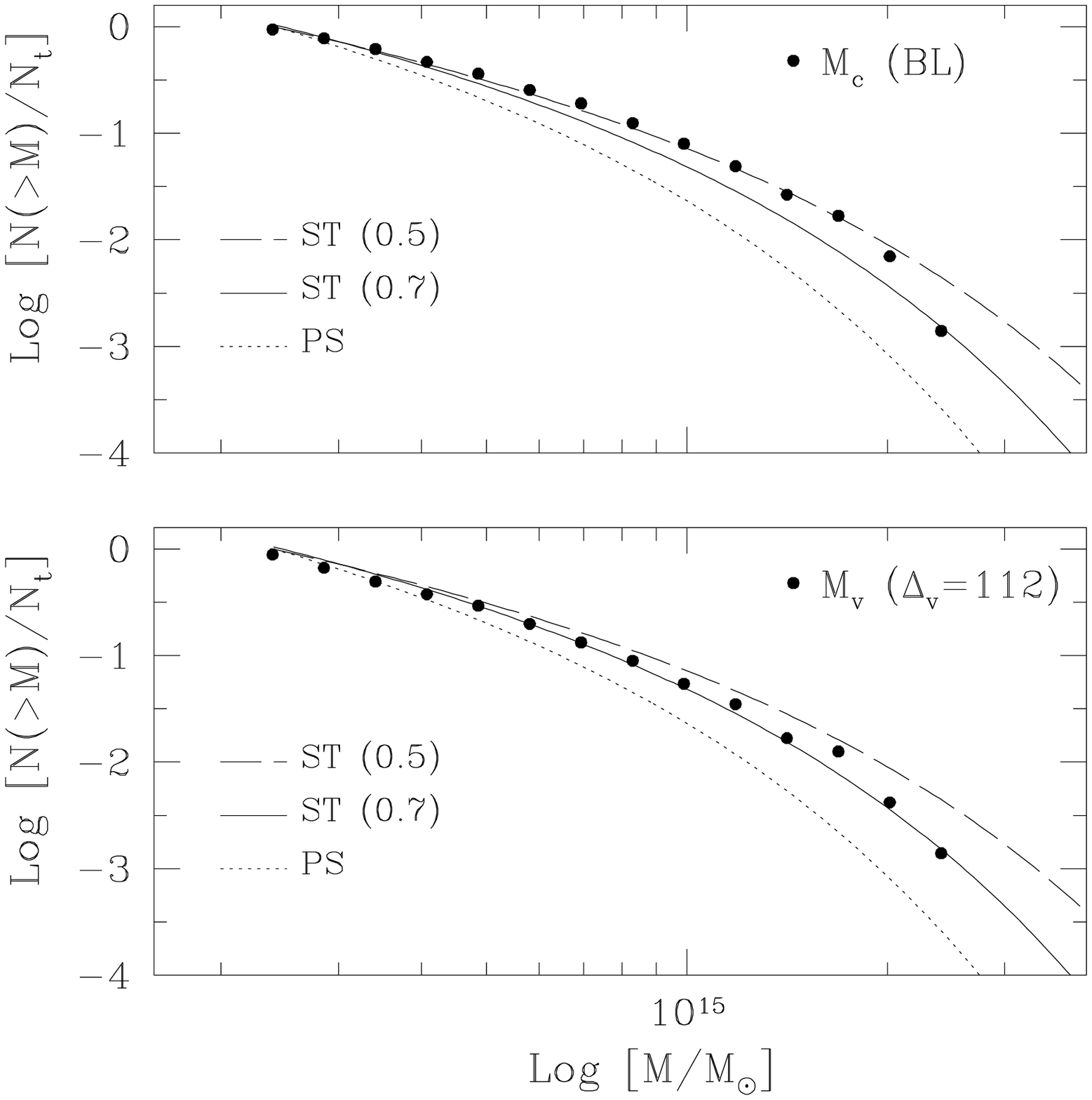}}}
\caption{Comparison between mass functions obtained using BL
or virial quantities ($\Lambda$CDM).
}
\label{Fig 14}
\end{figure}

\begin{figure}
\centerline{\mbox{\epsfysize=7.0truecm\epsffile{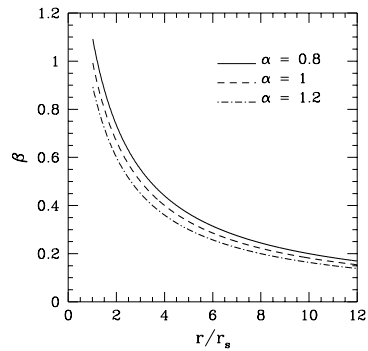}}}
\caption{$\beta$ yields the deviation from $3$ of the slope
of the halo profile in the region where the BL is expected to lay.
}
\label{Fig 15}
\end{figure}

\end{document}